\pdfoutput=1
\documentclass[conference]{IEEEtran}
\IEEEoverridecommandlockouts
\usepackage{algorithmic}
\usepackage[pdftex]{graphicx}
\usepackage{textcomp}
\usepackage{url}
\usepackage{balance}
\usepackage{paralist}
\usepackage{dblfloatfix}
\def\BibTeX{{\rm B\kern-.05em{\sc i\kern-.025em b}\kern-.08emT\kern-.1667em\lower.7ex\hbox{E}\kern-.125emX}}
\usepackage[table,xcdraw]{xcolor}
\usepackage[backend=bibtex,giveninits=true,hyperref=true,natbib=true,style=savetrees,sorting=none]{biblatex}
\usepackage[moderate,tracking=normal,paragraphs=tight,mathspacing=tight]{savetrees}
\usepackage{siunitx}
\usepackage{bm}
\usepackage{dsfont}
\usepackage{booktabs}
\usepackage{multicol}   
\usepackage{float}
\usepackage{csquotes}
\usepackage{hyperref}
\usepackage{placeins}
\usepackage{afterpage}

\definecolor{ForestGreen}{RGB}{34,139,34}\usepackage[ruled,vlined]{algorithm2e}
\usepackage{makecell}
    
\usepackage[english]{babel}
\usepackage{mathtools}
\usepackage{lipsum}
\usepackage{paralist}
\usepackage{enumitem}

\makeatletter
\let\MYcaption\@makecaption
\makeatother

\usepackage[font=footnotesize]{subcaption}

\makeatletter
\let\@makecaption\MYcaption
\makeatother
\usepackage{nicematrix}
\usepackage{colortbl}%
\usepackage{multirow}
\usepackage{footmisc} 
\usepackage{cleveref}

\newcommand{\smarthfill}{\hspace{1.2em minus 1.2em}}
\usepackage[left=1.62cm,right=1.62cm,top=1.9cm]{geometry}

\usepackage{tabularx}

\DeclareMathOperator*{\argmin}{arg\,min}
\newcounter{myenumi}
\newcommand{\R}{\ensuremath{\mathbb{R}}}
\makeatletter 
\newcommand{\linebreakand}{%
  \end{@IEEEauthorhalign}
  \hfill\mbox{}\par
  \mbox{}\hfill\begin{@IEEEauthorhalign}
}
\makeatother 

\usepackage[utf8]{inputenc}
\usepackage[T1]{fontenc}
\usepackage{hyphenat}
\usepackage{xspace}
\usepackage{amsmath}
\usepackage{amsfonts}
\usepackage{hyperref}
\usepackage{url}
\usepackage{booktabs}
\usepackage{multirow}
\usepackage{makecell}
\usepackage{minibox}
\usepackage{bbm}
\usepackage{graphicx}
\usepackage{balance}
\usepackage{mathtools}
\usepackage{color}
\usepackage{marvosym}
\usepackage{ifthen}
\usepackage{textcomp}
\usepackage{enumitem}
\usepackage{verbatim}
\usepackage{numprint}
\usepackage{balance}

\usepackage{amsthm}
\theoremstyle{plain}

\newcommand{\chatoDisplayMode}[1]{#1}

\definecolor{MyRed}{rgb}{0.6,0.0,0.0} 
\definecolor{MyBlack}{rgb}{0.1,0.1,0.1} 
\newcommand{\inred}[1]{{\color{MyRed}\sf\textbf{\textsc{#1}}}}
\newcommand{\frameit}[2]{
  \begin{center}
  {\color{MyRed}
  \framebox[.9\columnwidth][l]{
    \begin{minipage}{.85\columnwidth}
    \inred{#1}: {\sf\color{MyBlack}#2}
    \end{minipage}
  }\\
  }
  \end{center}
}

\newcommand{\note}[2][]{\chatoDisplayMode{\def\@tmpsig{#1}\frameit{{\Pointinghand} Note}{#2\ifx \@tmpsig \@empty \else \mbox{ --\em #1}\fi}}}
\newcommand{\todo}[2][]{\chatoDisplayMode{\def\@tmpsig{#1}\frameit{{\Writinghand} To-do}{#2\ifx \@tmpsig \@empty \else \mbox{ --\em #1}\fi}}}

\begin{document}

\title{SANDWICH: Towards an Offline, Differentiable, Fully-Trainable Wireless Neural Ray-Tracing Surrogate\\
}

\author{
    \IEEEauthorblockN{Yifei Jin}
    \IEEEauthorblockA{
        \textit{KTH Royal Institute of Tech. \& Ericsson AB} \\
        Stockholm, Sweden \\
        \url{yifeij@kth.se}
    }
\and
\IEEEauthorblockN{Ali Maatouk}
\IEEEauthorblockA{\textit{Yale University} \\
New Haven, USA  \\
\url{ali.maatouk@yale.edu}}
\and
\IEEEauthorblockN{Sarunas Girdzijauskas}
\IEEEauthorblockA{\textit{KTH Royal Institute of Tech.} \\
Stockholm, Sweden \\
\url{sarunasg@kth.se}}
\and
\linebreakand 
\IEEEauthorblockN{Shugong Xu$^1$~\thanks{$^{1}$The author was with the Shanghai University, Shanghai, China. He is now with the Xi'an Jiaotong-Liverpool University, Suzhou, China.}}
\IEEEauthorblockA{\textit{Xi'an Jiaotong-Liverpool University \& Shanghai University} \\
Suzhou \& Shanghai, China\\
\url{shugong@xjtlu.edu.cn}}
\and
\IEEEauthorblockN{Leandros Tassiulas}
\IEEEauthorblockA{\textit{Yale University} \\
New Haven, USA  \\
\url{leandros.tassiulas@yale.edu}}
\and
\IEEEauthorblockN{Rex Ying}
\IEEEauthorblockA{\textit{Yale University} \\
New Haven, USA  \\
\url{rex.ying@yale.edu}}
}

\maketitle
\setcounter{footnote}{2}
\begin{abstract}
Wireless ray-tracing (RT) is emerging as a key tool for three-dimensional (3D) wireless channel modeling, driven by advances in graphical rendering. Current approaches struggle to accurately model beyond 5G (B5G) network signaling, which often operates at higher frequencies and is more susceptible to environmental conditions and changes. Existing online learning solutions require real-time interaction with radio environment during training, which is both costly and incompatible with GPU-based processing. In response, we propose a novel approach that redefines ray trajectory generation as a sequential decision-making problem, solved with the proposed Scene-Aware Neural Decision Wireless Channel Raytracing Hierarchy (SANDWICH) approach. The SANDWICH approach leverages a decision transformer to jointly learn the optical, physical, and signal properties within each designated environment in a fully differentiable approach, which can be trained entirely on GPUs.
SANDWICH offers superior performance compared to existing online learning methods, and outperforms the baseline by $4e^{-2}$ \unit{\radian} in RT accuracy. Furthermore, channel gain estimation w.r.t predicted trajectory only fades $0.5$ \unit{\decibel} away from using ground truth wireless RT result for channel gain estimation.\textsuperscript{2}
\end{abstract}

\begin{IEEEkeywords}
Wireless Raytracing, RF Sensing, Channel Modeling, Channel Generation
\end{IEEEkeywords}
\footnotetext{This work was partially supported by the Wallenberg AI, Autonomous Systems and
Software Program (WASP) funded by the Knut and Alice Wallenberg Foundation. The implementation is available at \url{github.com/bluelancer/SANDWICH}}
\section{Introduction}
\label{sec:intro}
\begin{figure}
    \centering    
    \includegraphics[width=0.8\linewidth]{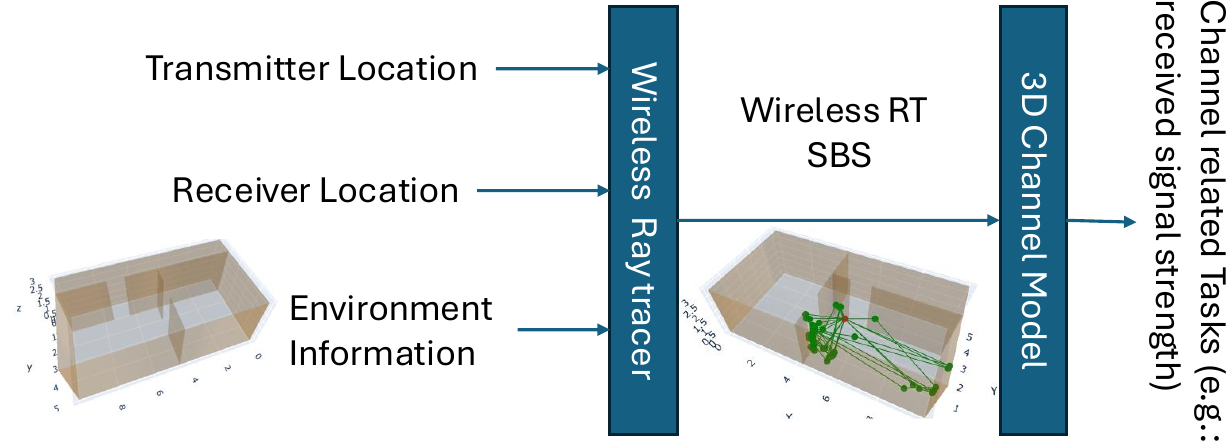}
    \caption{Workflow of Genuine Wireless Raytracer: Wireless Raytracer takes the input of Tx, Rx location, and radio environment information, and outputs SBS to be further utilized on wireless channel-related tasks.}
    \label{fig:rt}
\end{figure}

In the B5G/6G era, high-fidelity wireless channel modeling has become indispensable for advancing wireless systems. 
Recent studies in Joint Communication And Sensing (JCAS)~\cite{feng2024d}, beam management~\cite{papaioannou2022integrated}, and positioning~\cite{zhu2024physics} have underscored the necessity of accurately modeling wireless signal interactions within radio environments.
Supported by computational Electro-Magnetic (EM) theory and Geometrical Optics (GO). Recent research~\cite{hoydis2023sionna,amiot2013pylayers,choi2023withray,orekondy2023winert,tarneberg2023towards,zhang2024wisegrt,geok2018comprehensive,hoydis2023learning} introduces wireless RayTracing (RT) to capture these interactions through a Shoot-and-Bounce Sequence (SBS)~\cite{18706}. SBS is widely used in ray-tracing where rays are ``shot'' from a source and traced as they interact with surfaces through reflection, diffraction, and penetration. 
Wireless RT effectively models the EM wave propagation and scattering in a complex environment~\footnote{In this paper we mainly refer to indoor scenario.}, while adhering to GO and the Uniform Theory of Diffraction (UTD)~\cite{hoydis2023sionna}. 
Despite its potential, wireless RT involves computationally expensive raytracer~\cite{rtem}, which calculates a set of SBS relevant for wireless reception, from infinite GO \& UTD-feasible rays. In the data-driven channel modeling approach, such SBS set is a robust prior for a range of channel-related applications, as illustrated in Fig.~\ref{fig:rt}. We interchangeably refer to "wireless RT ray trajectory" and SBS throughout this work.
\subsection{Wireless RT Surrogate}
Recently, research has been exploring neural methods as wireless RT surrogates~\cite{orekondy2023winert,hoydis2023sionna,instant_rm} to estimate SBS, aiming to reduce the computational costs of traditional wireless ray tracers with scalability towards new Tx, Rx location. However, 
wireless RT is typically formulated as a sparsely-supervised and non-differentiable learning objective with Out-Of-Distribution (OOD) samples, for the following three reasons:
\begin{inparaenum}
    \item The indeterminate length of SBS and its interactions with the environment before its receipt by Rx;
    \item Due to the nature of the wireless signal, wireless RT SBS is not differentiable on the 2D plane of the environment.
    \item Similar to a genuine raytracer, the surrogate should be capable of handling unseen transmitter (Tx) \& receiver (Rx) locations in the 3D space.
\end{inparaenum}

Several approaches have been proposed in the literature to address such challenges. For example, \citet{orekondy2023winert} introduced WINERT, which mimics the output of a wireless ray tracer at each interaction with a greedy policy, using temporal difference (TD) learning.

While being capable of partially replacing a genuine raytracer with time \& computational superiority, existing methods are hindered by several significant limitations: \begin{inparaenum} 
\item \textbf{Online Supervision}: Dependence on real-time feedback from radio environment modeling and ray-tracer in training loop; 
\item \textbf{Vectorization}: Inability to process batches of rays efficiently, coupled with GPU incompatibility;
\item \textbf{Differentiability}: The existing method adapts the ray tracer to feed the state space to determine the next action, segmenting the radio environment into several discontinuous state spaces and neglecting their sequential relation, which leads to exploding gradients during training.
\end{inparaenum}

\begin{figure}[t]
    \centering
    \includegraphics[width=0.48\textwidth]{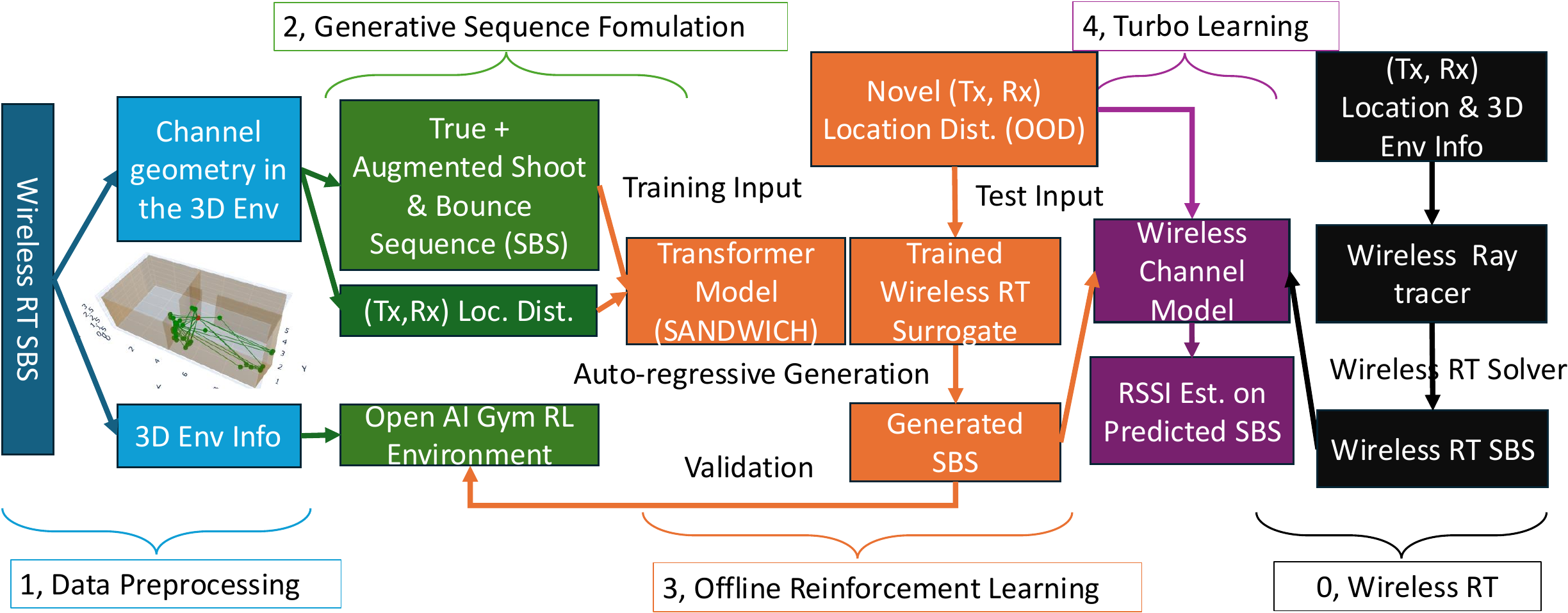}
    \caption{Schematic Representation of SANDWICH: $1$, Segment wireless RT and 3D environment. $2$, Sequentialize RT result into SBS. $3$, Train neural surrogate \& apply auto-regressive generation. $4$, Apply generated RT to the wireless channel model. Such workflow serves as a neural surrogate of $0$, a wireless RT process.}
    \label{fig:schematicRep}
\end{figure}

\subsection{Proposed Method} 
This paper addresses wireless RT considering Tx/Rx locations and signal-surface interaction in indoor environments with surface texture information. To capture ``state space continuity'', we reformulate the wireless RT as a sequence generation problem~\cite{radford2018improving}. Inspired by the recent success of offline Reinforcement Learning (RL) algorithms like Decision Transformer(DT)~\cite{chen2021decision}, we propose a novel Scene-Aware Neural Decision WIreless Channel raytracing Hierarchy (SANDWICH) model. The method models ``state space continuity'' and allows learning from a limited set of collected RT results without assumptions on the radio environment or supervision density. Specifically, SBS representing a wireless ray can be generated similarly to token sequences, utilizing the transformer's~\cite {vaswani2017attention} capability of learning relational attention within the token sequence.
We house the sequence auto-regressive generation task within an offline RL scheme, enforcing the model to internalize the radio environment as a constraint during SBS generation, to unhook the requirement for online supervision from well-parameterized radio environment in State-Of-The-Arts (SOTA).

We illustrate a schematic representation of the proposed solution in Fig.~\ref{fig:schematicRep}, including four major steps:
\begin{inparaenum}
    \item Segment raw data into 3D environment and channel information. The ray's trajectory is also transformed into true \& augmented SBS for transformer models.
    \item Create tailored \textsc{OpenAI Gym} environment for test-time verification with OOD Tx \& Rx locations. 
    \item Train the proposed model and generate SBS on novel (Tx, Rx) distribution.
    \item Besides the generated wireless RT geometrical accuracy, the generated SBS are utilized as priors for channel model, to enhance performance in downstream tasks within a Turbo Learning~\cite{zhao2023nerf2} framework.
\end{inparaenum}

Our method leverages a customized DT~\cite{chen2021decision} to integrate sparse supervision through expected returns through a novel sequential decision-making formulation for Markov Decision Process (MDP), thereby decoupling the need for continuous online supervision from the environment. Additionally, we propose a data augmentation technique based on Fibonacci sphere~\cite{rogne2022raytracing} to generate stochastic trajectories that enhance generalization of DT on OOD test samples, alongside state supervision to enforce environment awareness.

\subsection{Our Contribution} 
\begin{inparaenum}
\item We propose an end-to-end
GPU-trainable, fully differentiable, and vectorizable SANDWICH framework for wireless RT.
\item In addition, we propose to augment the training rays by sampling from Fibonacci Spheres, that proven to strengthen the generalization performance.
\item we also propose to distill the optical properties of materials to the model by predicting the interaction type.
\item Our approach outperforms the online learning solution by $4e^{-2}$ \unit{\radian} in RT accuracy and only fades $0.5$ \unit{\decibel} away from GT-ray-powered channel gain estimation, while outperforming all non-RT based baselines with generated wireless RT results.
\end{inparaenum}

\section{Related Work}
\label{sec:related}

\subsection{Wireless 3D Channel Modeling}
Wireless RT is a specialized subfield within wireless channel estimation, particularly relevant in 3D environments where millimeter-wave (mm-wave) propagation can be statistically described as~\cite{samimi20163}
\begin{equation}
\label{equ:channel}
    h(t,\Theta,\Phi)=\small\sum_{r_k \in \{r_k\}}a_k(t) \delta(t-\tau_k(t)) 
    \delta(\Theta-\Theta_k(t))\\
    \delta(\Phi-\Phi_k(t))
\end{equation}
In~\eqref{equ:channel}, a wireless channel is represented by a set of rays (denoted as $\{r_k\}:= \{r_1, \cdots, r_k\}$, with cardinality $|\{r_k\}| = K$). Each ray $r_k$ is characterized by its signal gain $a_k(t)$, one-way delay $\tau_k(t)$, 
and phase $\Theta_k(t), \Phi_k(t)$ for arrival and departure angle, respectively, expressed using Dirac function $\delta$. 
These parameters summarize the property of SBS, which includes a series of ray-surface interactions, in the radio environment $F$. 

\subsection{Heuristic Wireless RT} 
Recent research has focused on wireless RT estimation in a 3D environment and application to physics-informed channel modeling~\cite{zhu2024physics}. \citet{geok2018comprehensive} offers a comprehensive review of the heuristics method and discusses such methods' complexity and generalization limitations. \cite{amiot2013pylayers, remcomRemcomElectromagnetic} offers heuristics-method-driven simulator in 3D environment, which has been proved by \citet{orekondy2023winert} to be time and computational expensive. \citet{hoydis2023sionna, choi2023withray} proposed a SBS-based simulator that greatly accelerates the simulation speed through highly efficient Monte-Carlo rays integration while dependent on online supervision from the environment through a designated ray-tracing engine. 
SANDWICH builds upon insights from heuristic from~\citet{geok2018comprehensive,hoydis2023sionna}, with the concepts of SBS and early-step-focused-rewarding to mitigate bias propagation within a generative portfolio~\cite{ladhak2023pre}. 

\begin{figure*}[thb]\captionsetup[subfigure]{font=small}
\centering
    \begin{subfigure}[t]{0.35\textwidth}
        \includegraphics[width=\linewidth]{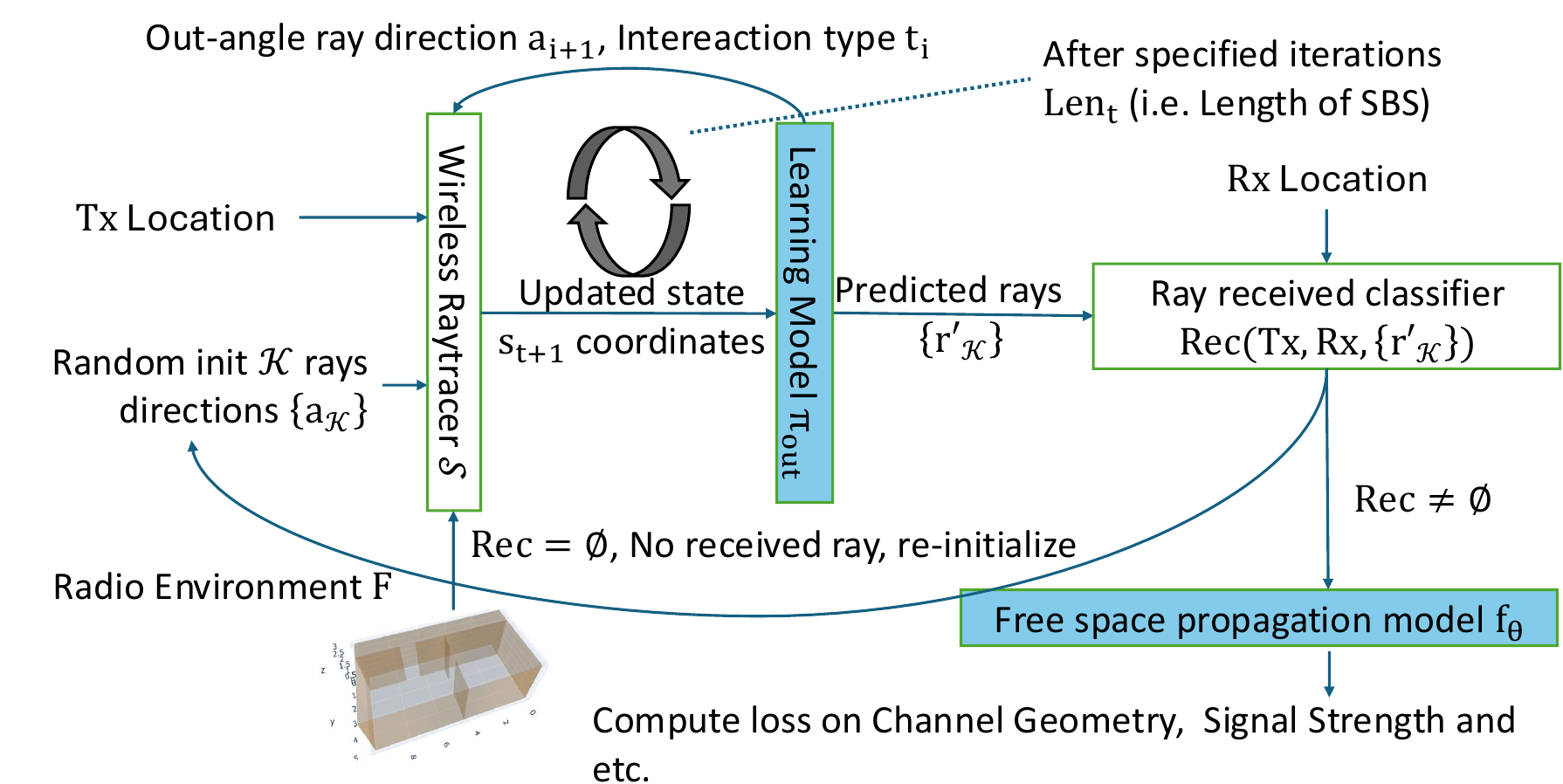}       \captionsetup{justification=raggedright,singlelinecheck=false}
        \caption{Schematic Representation of Problem Formulation in \cite{orekondy2023winert}: 
        \begin{inparaenum}
            \item[(a,1)]\label{form:prevPFa1} Given Tx location and $\mathcal{K}$ randomly initialize ray direction $\{a_\mathcal{K}\}$, $\mathcal{S}$ outputs the state $s$ regarding the intersection of $\{a_\mathcal{K}\}$ and $F$.
            \item[(a,2)]\label{form:prevPFa2} The greedy policy $\pi_{out}$ was trained by iteratively collecting next-state $s_{t+1}$ from $\mathcal{S}$ to suggest current type and next direction.
            \item[(a,3)]\label{form:prevPFa3} Once the designated length $\text{Len}_k$ is reached, a logic-based ray receipt classifier $Rec$ will filter the predicted SBS proximate to Rx location. The training is constrained to be batch size $1$ to ensure greediness.
        \end{inparaenum}}
        \label{fig:prevPF}
    \end{subfigure}%
    \smarthfill
    \begin{subfigure}[t]{0.62\textwidth}
        \includegraphics[width=\linewidth]{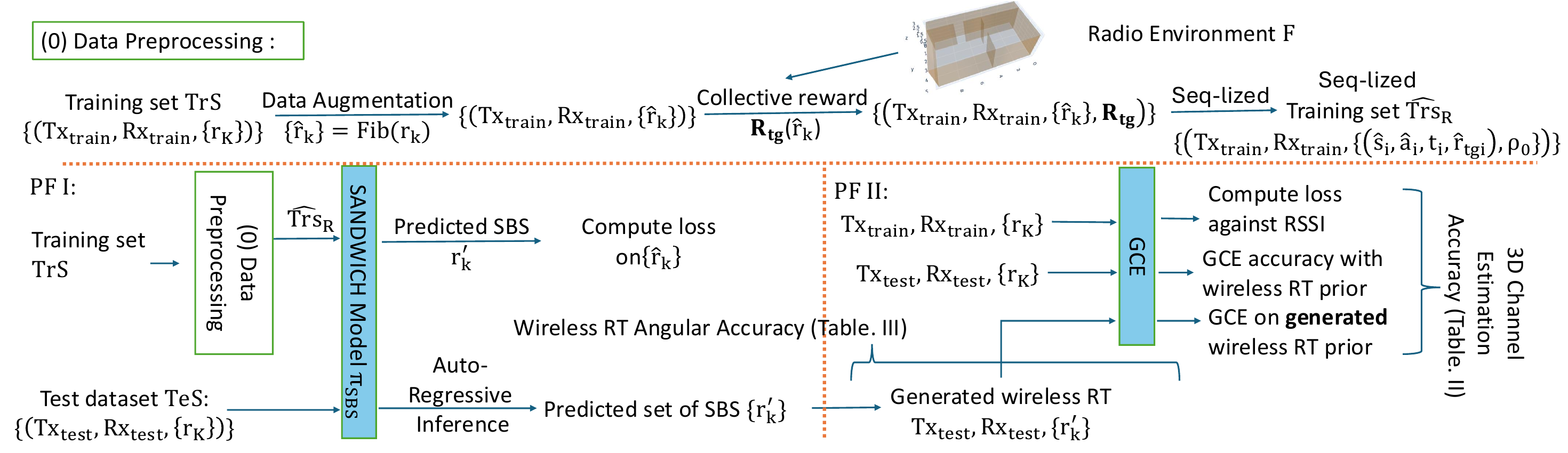}
        \captionsetup{width=0.8\linewidth,justification=raggedright,singlelinecheck=false}
        \caption{Schematic Representation of Proposed Problem Formulation: 
        \begin{inparaenum}
            \item[(b,1)]\label{form:newPFb1} A collected set of SBS is augmented, attributed with collective reward, and sequentialized to be adapted to sequential-decision-making formulation. 
            \item[(b,2)]\label{form:newPFb2} In \textbf{PF I}, the learning policy $\pi_{SBS}$ takes observation on the whole SBS, with a seq2seq training scheme.
            \item[(b,3)]\label{form:newPFb3} In test time, $\pi_{SBS}$ takes the input of OOD Tx \& Rx and is promoted to generate the sequence with maximum collective reward. The generated $\{r'_k\}$ will be used in \textbf{PF II}.
            \item[(b,4)]\label{form:newPFb4} To compare the usability of generated SBS, we compare the generated $\{r'_k\}$ and GT $\{r_k\}$ through a well-trained GCE and offer precision and error distribution analysis.
        \end{inparaenum}}
        \label{fig:updatePF}
    \end{subfigure}
    \caption{Comparison of Problem Formulations on Learning Wireless RT Surrogate: Each blue box denotes a trainable component and the white box denotes a non-trainable component.}
\label{fig:compare_PF}
\end{figure*}

\subsection{Neural Wireless RT} 
In the past year, researchers have been distilling the complex heuristic (or parts of it) into neural surrogates for real-time Wireless RT results~\cite{lai2024real,zhao2023nerf2,hoydis2023learning}, where \cite{lai2024real,zhao2023nerf2} adapt Neural Radiance Field (NeRF) to extract 3D mesh from RGB images while \cite{hoydis2023learning} incorporate an online learning formulation.
The closest work to ours is by \citet{orekondy2023winert}, where the authors proposed WINERT based on a TD-learning approach to distilling wireless RT. 
However, their method is not vectorizable, not GPU trainable, and supervised by online feedback from the radio environment. 
Recent work~\cite{hehn2024differentiable,chi2024rf} has avoided wireless RT but attempted to model RF signal strength distribution from a probabilistic perspective without SBS prior, while this goes beyond the scope of wireless RT surrogate and does not capture the full wireless propagation physics.
We consider SANDWICH shares the strength of \cite{hoydis2023learning,orekondy2023winert,knodt2021neural} through a novel generative formulation of wireless RT problem while being offline, fully differentiable, and vectorized-trainable with scene-awareness.
Instead of decoupling the SBS into densely populated rewards, our approach generates SBS in an auto-regressive generation manner, spanning the sparse-reward signal through the whole SBS trajectory.

\subsection{MDP on Wireless RT}
Given the locations of the Tx and Rx, denoted by $\text{loc}(Tx)$ and $\text{loc}(Rx)$ respectively, a set of rays $\{r_K\}$ are emitted from Tx and interacts with $F$ through a designated simulator $\mathcal{S}$ until received at Rx~\cite{amiot2013pylayers, remcomRemcomElectromagnetic}. Following the work in~\cite{orekondy2023winert}, the RT training formulation can be done through an MDP $(\mathbf{S}, \mathbf{A}, \mathbf{P})$, decomposing each ray in SBS set $r_k \in \{r_K\}$ into a sequence of state-action-interaction triplets $[(s_i,a_i,t_i)]$, state space $s_i \in \mathbf{S} \subset \R^3$, action space $a_i \in \mathbf{A} \subset SO(3)$, where $SO(3)$ denotes 3D rotation group, and transition dynamics $\mathbf{P}$. Each state $s_i$ represents the coordinate of a ray in its trajectory, while each action $a_i = (\phi,\theta) \in \mathbf{A}$ denotes the radian $\phi$ and azimuth $\theta$ out-angle relative to the previous hop $s_{i-1}$. The interaction $t_i$ at each state $s_i$ can be categorized into three types: reflection ($\mathcal{R}$), diffraction ($\mathcal{D}$), and penetration ($\mathcal{P}$) that attributes on $s_i$ as property of wireless signals. The transition function is defined as $\mathbf{P}(s_{i+1}|s_i, a_i) = \mathcal{S}(F, s_i, a_i)$ through $\mathcal{S}$ as collected experience in SBS. A logic-based receiver classifier $\text{Rec}(Rx, F, \{r_K\})$ determines if the rays in $\{r_K\}$ successfully reach the Rx. The objective is to learn a greedy policy $\pi_{out}$ that predicts identical ray set $\{r_K\}$ on new $\text{loc}(Tx)$, $\text{loc}(Rx)$ combined with $\mathcal{S}$ as transition dynamics $\mathbf{P}$, such formulation is outlined in Fig.~\ref{fig:prevPF}.

\section{Problem Formulation}
\label{sec:formulation}

\subsection{Challenge in MDP Formulation}
MDP formulation of wireless RT integrates several inherent challenges:
\begin{inparaenum}
    \item \textbf{Sparse Supervision/reward}: Reward signals are granted on receipt of each SBS, spanning across a sequence of decisions, leading to sparse supervision.
    \item \textbf{Non-differential Objective}: Wireless ray propagation is influenced not only by the locations and configurations of the Tx and Rx but also by the 3D structure and texture of the environment. Consequently, wireless RT results may not be differentiable in a 2D plane, emphasizing the need for a sequential representation-based learning approach.
    \item \textbf{OOD Samples in $\mathbf{S}$}: To demonstrate the generalization capabilities of the proposed model, the test environment includes novel Tx and Rx locations $\text{loc}(Tx)$, $\text{loc}(Rx)$, resulting in an extrapolated state space. This highlights the necessity of incorporating $F$ as an inductive bias.
\end{inparaenum}

\subsection{Limitation on Alleviation Approaches}
Although some SOTA methods~\cite{orekondy2023winert,hoydis2023sionna} can alleviate such challenges, however these alleviations themselves introduce new problems:
\begin{inparaenum} 
    \item \textbf{Densifying the reward signal on a per-action basis}: MDP formulation densifies such rewards by learning a greedy policy $\pi_{out}$ that predicts the next action $a_{i+1}$ and current type $t_i$ with barely current state $s_i$, and action $a_i$. 
    This often introduces heuristic and non-differentiable components $\mathcal{S}$ and $\text{Rec}$ in training time as non-trainable components.
    Moreover, ray length $\text{Len}_k$ has to be specified which leaks information during test time;
    \item \textbf{Excessive exploring $F$}: SOTA methods emits a large number of rays $\{r_\mathcal{K}\}$, where $|\{r_\mathcal{K}\}| = \mathcal{K} \gg K$, to explore the reward distribution in $F$ through online supervision from $\mathcal{S}$' feedback.
\end{inparaenum}
\textbf{We argue that the excessive computation is caused by the fact that the SBS ray generation process, is not physically a markov process.} 
Hence modeling it with an MDP necessitates excessive sampling and post-trajectory pruning with $Rec$. 
These approaches incur significant computational overhead, create incompatibility on GPU due to $\mathcal{S}$ involvement, and rely on online supervision.

In contrast, this paper formulates the wireless RT task as \textbf{an end-to-end learning model to retrieve the set of SBS for OOD Tx \& Rx location, acquired by a wireless raytracer}, which leads to an offline, fully GPU-trainable method (SANDWICH) that learns wireless RT through a generative scheme.

\begin{table}[htbp]
\caption{Table of Notations}
\label{table:notation}
\centering
\begin{tabularx}{\linewidth}{r X}
\toprule 
\multicolumn{2}{l}{\textbf{Wireless 3D Channel Modeling Notation:}}\\
$h(t, \Theta, \Phi)$ & Wireless channel response \\
$r_k$ & A single ray in the wireless RT result $\{r_k\}$ \\
$K$ & Cardinality of the set of rays $\{r_k\}$ for $h(t, \Theta, \Phi)$\\
$a_k(t)$, $\tau_k(t)$ & Signal gain, time delay of the ray $r_k$ at time $t$ \\
$\Theta_k(t), \Phi_k(t)$ & Angular parameters of the ray $r_k$ at time $t$ \\
$\delta(\cdot)$ & Dirac delta function \\
$loc(Tx), loc(Rx)$ & Locations of the transmitter and receiver \\
$F$ & Wireless radio environment \\
\midrule 
\multicolumn{2}{l}{\textbf{MDP with TD-Learning Formulation Notation:}}\\
$\{r_\mathcal{K}\}$ & A large number of ray attempts (i.e.$|\{r_\mathcal{K}\}| = \mathcal{K} \gg K$) for TD Learning \\
($\mathbf{S},\mathbf{A}$,$\mathbf{P}$) & State space, Action space and Transition dynamics\\
($s_i$,$a_i$,$t_i$) & State, action, interaction triplets of the ray at step $i$ \\
$P(s_{i+1}|s_i, a_i)$ & Transition function of moving to state $s_{i+1}$, realized by  designated ray-tracer/RT Simulator $\mathcal{S}(F, s_i, a_i)$\\
$\pi_{out}(s_i, a_i)$ & TD-Learning policy for predicting next ray direction $a_{i+1}$ and current interaction type $t_i$ \\
$R(s_i, a_i)$ & Reward function at state $s_i$ for action $a_i$ \\
$\rho_0$ & Initial action distribution \\
$\text{Len}_k$ & Length of the ray's shoot-and-bounce sequence (SBS) \\
$r_k$ & Full SBS trajectory of the ray $k$: $[(s_i, a_i, t_i)]$ \\
$Rec(Rx, F, \{r_k\})$ & Receiver classifier for determining valid rays \\
\midrule
\multicolumn{2}{l}{\textbf{Sequential Decision-Making Formulation Additional Notation:}}\\
$\pi_\text{SBS}(Tx, Rx, \rho_0)$ & Neural sequential policy for generating SBS \\
$\{r'_k\}$ & Generated ray set by SANDWICH\\
$\hat{r_k} = \mathbf{Fib}(r_k)$ & Augmented SBS $\hat{r_k}$ using Fibonacci sphere, from each sample $r_k$ in training set. \\
$\sigma_\phi, \sigma_\theta$ & Variance in Fibonacci augmentation \\
$\bm{R}_{tg}(r'_k)$ & Collective reward for the generated ray $r'_k$, $r_{tgi} \in \bm{R}_{tg}(r'_k)$ for each step $i \in [L_k]$, also applied on $\hat{r_k}$.\\
$R_{\text{SBS}}(r'_k, \{r_k\})$ & Set-based fitness for $r'_k$ against $\{r_k\}$ to parameterize $\bm{R}_{tg}(r'_k)$, consists of $\textsc{L}_{\text{ray}}$ and $P_\angle$, also applied on $\hat{r_k}$ against $\{r_k\}$ for augmented sample's fitness.\\
$\textsc{L}_t(r'_k, F)$ & Training loss of SANDWICH, parameterized by $\textsc{L}_{\text{ray}}$, $\textsc{L}_{\text[type]}$ and weighting $\alpha$ \\
$\textsc{L}_{\text{ray}}(r_k, r'_k)$ & Set-based loss on SBS geometrical fitness\\
$\textsc{L}_{\text[type]}(t_i, t'_i)$ & Cross-entropy loss on $t_i$\\
$P_\angle(r'_k, F)$ &  Geometric correctness penalty, consists of $\textsc{bu}, \textsc{ob}, \textsc{te}$\\
$\textsc{bu}(\{(a'_j, t_j)\})$ & Geometric consistency penalty at step $j$ \\
$\textsc{ob}(r'_k, F)$ & Outbound penalty function for $r'_k$ regarding $F$\\
$\textsc{te}(r'_k,Rx)$ & Terminal error for $r'_k$\\
\bottomrule 
\end{tabularx}
\end{table}

\subsection{Proposed Problem Formulation}
In this work, we introduce two novel problem formulations (PF) following previous acronyms:
\textbf{PF I:} Given MDP $(\mathbf{S}, \mathbf{A}, \mathbf{P}, \bm{R}_{tg}, \rho_0)$, we consider each SBS $r_k$ to be summarized as $[(s_i,a_i,t_i)]$, with length $\text{Len}_k$ and deterministic transition $\mathbf{P}$ from collected wireless RT results. $\bm{R}_{tg}$ is a set-based reward function evaluates a predicted SBS $r'_k$ against GT set $\{r_k\}$, and $\rho_0$ denotes the initial action distribution.
The objective is to find a policy $\pi_\text{SBS}$ that generates SBS $r'_k$, maximizing the expected return $E(\bm{R}_{tg}(r'_k))$.

Additionally, to validate the utility of fully-generated wireless RT sequences $\{r'_k\}$, we incorporate a secondary problem formulation through \textit{Turbo Learning}:
\textbf{PF II}: We train a ``Golden Channel Estimator'' (GCE) $h'_G$ using data-driven approach~\cite{sitzmann2020implicit,} to simulate $h(t, \Theta, \Phi)$ in \eqref{equ:channel}. This is done by summarizing $\{r_k\}$, $F$, $Rx$, $Tx$, and their respective locations $\text{loc}(Tx)$ and $\text{loc}(Rx)$ all as inputs to the estimator. We conduct a hyperparameter search to optimize the GCE for its best possible channel estimation. We then compare the performance and error (KL-divergence) with predicted $\{r'_k\}$ against the GT $\{r_k\}$ on channel estimation task using GCE. Both PFs are tested against novel distribution of (Tx, Rx) on extrapolated feature space to prove its generalization performance. We summarize the differences between our problem formulation and previous works in Fig.~\ref{fig:compare_PF}, and summarize the defined notation in Table.~\ref{table:notation}.

\section{Methodology}
\label{sec:method}
\begin{table*}[htb]
\centering
\caption{Comparison of MAE and KL Divergence between Models and GT/GCE across the testsets}
\begin{tabular}{lS[table-format=3.2]S[table-format=3.2]S[table-format=3.2]S[table-format=3.2]S[table-format=3.2]S[table-format=3.2]S[table-format=3.2]S[table-format=3.2]S[table-format=3.2]} 
\toprule
\textbf{Metric (\unit{\decibel})}                      & \multicolumn{3}{c}{\textbf{MAE(Model, GT)}} & \multicolumn{3}{c}{\textbf{MAE(GCE, Model)}} & \multicolumn{3}{c}{\textbf{KL\_div(pdf(GCE)-pdf(Model))}} \\ 
\midrule
\textbf{On Test set:}           & {Checkerboard}  & {Genz} & {GenDiag} & {Checkerboard}  & {Genz} & {GenDiag} & {Checkerboard}  & {Genz} & {GenDiag}  \\ 
\cmidrule(lr){2-4}\cmidrule(lr){5-7}\cmidrule(lr){8-10}
GCE (Topline) & 3.422  & 3.181  & 3.039  & 0  & 0  & 0  & 0  & 0  & 0 \\ 
SANDWICH (Proposed)                & \bfseries 3.966 & \bfseries 3.804  & \bfseries 3.689 & \bfseries 1.494  & \bfseries1.640  & \bfseries 1.422  & \bfseries 0.0410   & \bfseries 0.0534  & \bfseries 0.0317 \\ 
MLP(Reproduced)          & 5.337  & 5.846  & 4.729  & 3.760  & 4.291  & 3.496  & 0.2827  & 0.6003  & 0.1913 \\  
KNN(Reproduced)~\ref{knn:1}          & 11.21  & 11.04  & 4.412  & 9.856  & 9.939  & 4.963  & 0.1447  & 0.1565  & 0.1468 \\ 
KNN(Reproduced)~\ref{knn:2}          & 10.71  & 10.59  & 4.180  & 9.446  & 9.542  & 4.822  & 0.1431  & 0.1516  & 0.1508 \\ 

\bottomrule
\end{tabular}
\label{tab:mae_kl_comparison}
\end{table*}
\begin{figure}[t!]
    \centering    
    \includegraphics[width=0.48\textwidth]{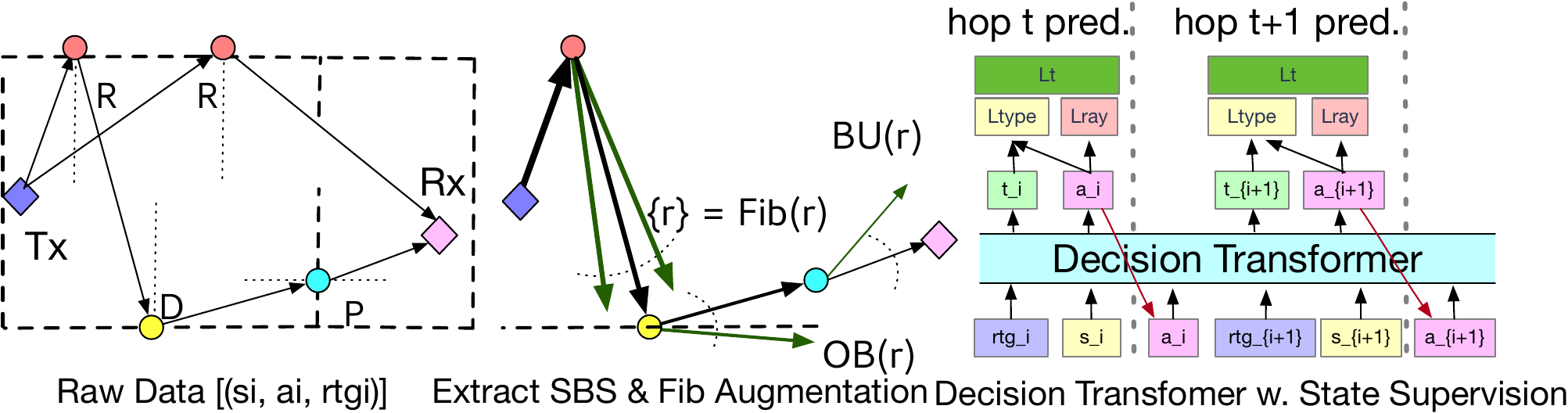}
    \caption{Component in SANDWICH Training Pipeline (i.e. \textbf{PF I} in Fig.~\ref{fig:updatePF}):
    $i)$ SBS Sequentialization: Given a set of SBS from Tx to Rx in wireless RT result, each SBS is parsed into a decision sequence, with interaction type (noted as differently colored shapes) attributed on each step. 
    $ii)$ Per-SBS Fib Augmentation: For each SBS, we applied $\mathbf{Fib}(r_k)$ to augment a diverse quality of SBS, and compute collective reward $\bm{R}_{tg}$ as their quality metric.
    $iii)$ Train SANDWICH through the augmented \& shuffled SBS: apply seq2seq training on the input sequence $\{s_i,a_i,r_{tgi}\}$, against output sequence $\{t_i,a_{t+1}\}$ regards to loss function $\textsc{L}_{t}$.}
    \label{fig:method}
\end{figure}
The key idea is to \textbf{keep track of the historical states in each SBS to improve sampling efficiency when collecting experience from radio environment}. This is critical to handle the sparse rewards of wireless RT within an offline RL framework, as the expected return is granted for the whole SBS, outlined in \textbf{PF I}. 

Modeling SBS as a sequence aligns with optical ray tracing techniques like \textit{path tracing}~\cite{carrpath}, which \citet{clarberg2022keynote} demonstrated to be more precise than per-hop ray-surface interaction modeling, albeit computationally impractical for real-time rendering. Wireless RT strikes a balance, as noted by \citet{kajiya1986rendering}: while path-tracing a finite number of rays is computationally expensive, it remains feasible and sufficient for downstream tasks (e.g., multi-path modeling), enhancing ray rendering precision. Such insight diverges from greedy policy's state embedding~\cite{orekondy2023winert}, (Tx,Rx)-conditioned embedding~\cite{tancik2020fourier, sitzmann2020implicit}, or upsampling methods used to densify existing RT results~\cite{sobehy2020csi} in previous works.

\subsection{Decision Transformer}
DT~\cite{chen2021decision} is promising for handling sequential decision-making by referencing all prior decisions within a sequence. 
As formulated in \textbf{PF I}, each SBS is represented as a token sequence, where each step is defined by a triplet of its state, action, and return $(s_i, a_i, r_{tgi})\in (\mathbf{S}, \mathbf{A}, \bm{R}_{tg})$.  
For a fair comparison, we input the sequence of all preceding steps $[(s_i, a_i, r_{tgi})]_{|i}$ up to hop $i$, masking the current action $a_i$ to predict the current hop’s action and interaction type $t_i$, as \cite{orekondy2023winert} did.
At test time, DT models leverage the expected return $E(\bm{R}_{tg}(r'_k))$ to perform auto-regressive generation for OOD test inputs, defined by $\text{loc}(Tx)$ and $\text{loc}(Rx)$, against generated ray $r'_k$. The return function $\bm{R}_{tg}$ represents a cumulative reward for future steps, extending beyond the greedy reward used in MDP formulation (i.e.: illustrated in ~\ref{form:prevPFa1}). $\bm{R}_{tg}$ also incorporates a near-differentiable RT precision metric to guarantee optimal generation precision. This approach frames wireless RT as a differentiable, end-to-end sequential decision-making problem, as illustrated in Fig.~\ref{fig:method}.

\subsection{Fib Augmentation}
DT shares a suboptimal generalization performance as per~\cite{kimdynamics, ma2023rethinking}, especially in SBS testset including OOD Tx, Rx locations. Moreover, wireless RT SBS only includes positive samples, without null samples to shape the action space. Inspired by~\citet{rogne2022raytracing}, we introduce a data augmentation scheme, $\mathbf{Fib}({r_k})$, leveraging the Fibonacci sphere to improve sampling efficiency from radio environment with non-trivial exploration dynamics. 
We generate a batch of normal Fibonacci spheres for each $a_j$ and sampled a set of augmented ray $\mathbf{Fib}(r_k) = \{[(\hat{a_j}, t_j)]_k\}$ by introducing tunable variances $(\sigma_\phi,\sigma_\theta)$ around $(\phi,\theta)$. The set of augmented ray is shuffled together with SBS GT to form a training set of SBS:
\begin{equation}
\label{equ:aug}
\{\hat{r_k}\} := \{r_k\}\cup \big\{\bigcup_{r_k\ \in \{r_k\}}\mathbf{Fib}(r_k)\big\} 
\end{equation}

\subsection{Reward Shaping}
To offer a near-differentiable reward measure $\bm{R}_{tg}$~\cite{chen2021decision}, we propose a novel fitness function $R_{\text{SBS}}(\hat{r_k}, \{r_k\})$ to formulate the collective reward $\bm{R}_{tg}(\hat{r_k})$. We expect $R_{\text{SBS}}$ to measure $\{\hat{r_k}\}$ from both RT precision and geometry constraints in radio environment:
\begin{equation}
\label{equ:RSBS}
    \begin{split}
    \bm{R}_{tg}(\hat{r_k}) := & R_{\text{SBS}}(\hat{r_k}, \{r_k\}) \\
    =& -\argmin_{r_k\in \{r_k\}}\left[\textsc{L}_{\text{ray}}(r_k,\hat{r_k}) -P_\angle(\hat{r_k}, F)\right]  \\
    \end{split}
\end{equation}
 In \eqref{equ:RSBS} $, \textsc{L}_{\text{ray}}$ is an angular set-based loss that supervises the RT precision scale:
\begin{equation}
\label{equ:Lray}
    \textsc{L}_{\text{ray}}(r_k,\hat{r_k}) =  \textstyle \sum_{j \in [L_k]}w_j \log(|{a_j} - \hat{a_j}|), \quad a_j\in r_k.
\end{equation}
In \eqref{equ:Lray}, $w_j$ is a heuristic reverse sequential weight to encourage the transformer to learn the Markovian properties of the sequence. Unlike an increasing weighted reward in Vanilia DT, this decreasing weighted reward emphasizes supervision on the closest match to the GT SBS during the early generation steps. This approach helps avoid a trivial solution that could arise between two optimal SBS. On the other hand, $P_\angle$ supervises the geometry constraints:
\begin{equation}
\label{equ:pangle}
    P_\angle(\hat{r_k}, F) = \textsc{ob}(\hat{r_k}, F)+\textstyle \sum_{j \in |\hat{r_k}|}\textsc{bu}\left(\hat{a_j}, t_j\right)+\textsc{te}\left(\hat{r_k},Rx\right).
\end{equation}
In \eqref{equ:pangle}, $\textsc{ob}(\hat{r_k}, F)$ penalizes any outbound attempts by $\hat{r_k}$ from $F$. The term $\textsc{te}(\hat{r_k},Rx)$ introduces a penalty $e^d$, where $d \geq d_0$, if $\hat{r_k}$ misses $\text{loc}(Rx)$ by a tunable distance $d_0$. Lastly, $\textsc{bu}(\hat{a_j}, t_j)$ ensures geometric consistency between $(\hat{a_j}, t_j)$
\begin{equation}
\label{equ:BU}
  \begin{split}
  \textsc{bu}\left(\hat{a_j}, t_j\right) =
    \begin{cases}
      \argmin_{ax}\left[|\mathds{1}_{ax}\hat{\bm{a}_j}|-|\mathds{1}_{ax}\hat{\bm{a}_{j-1}}|\right] \quad \text{if } t_j=\mathcal{R} \\
      -10 \quad \text{if } t_j = \mathcal{P}, \hat{a_j} \neq \hat{a_{j+1}} \\
      0 \quad \text{otherwise.}
    \end{cases}
  \end{split}
\end{equation}
In \eqref{equ:BU}, $\hat{\bm{a}_j}$ is the directional vector defined by $\hat{a_j}$, while $\mathds{1}_{ax}$ denotes a set of unit vectors along the 3D axes. 

$R_{\text{SBS}}$ aligns the expected return $\bm{R}_{tg}$ from DT with wireless RT precision while addressing geometry constraints.
Such correspondence fitness is inspired by the adversarial clustering method described in~\cite{paster2022you}.
The holistic metric of SBS augmentation samples enables DT to learn a sequential action space for SBS, decouple the requirement of collecting per-action reward from simulators (in contrast to~\cite{orekondy2023winert, wang2024graph}), and empowers vectorization and GPU-training.

\subsection{Optimization} 
Finally, we formulate the training loss $\textsc{L}_{t}(r'_k, F)$ of SANDWICH against each predicted SBS $r'_k$:
\begin{equation}
    \openup 1\jot
    \begin{split}
    &\textsc{L}_{t}(r'_k, F)=\argmin_{r_k \in\{r_k\}} \left[\alpha \textsc{L}_{\text{ray}}(r_k,r'_k) + \sum_{i \in [L_k]}\textsc{L}_{\text[type]}(t_i,t'_i) \right]  \\ 
    &\textsc{L}_{\text[type]}(t_i, t'_i)=\textsc{CrossEntropy}(t'_i,t_i)
    \end{split}
\end{equation}
where $\alpha$ is a hyperparameter. Additionally, we incorporate state supervision $\textsc{L}_{\text[type]}$ into the backbone DT to enforce surface texture during ray interactions, inspired by DADT.~\cite{kimdynamics}.

We show the effectiveness of novel components, $\mathbf{Fib}(r_k)$ and $\textsc{L}_{\text[type]}$, contribute SANDWICH performance in Sec.~\ref{sec:exp}. 
The aforementioned building blocks are housed in an end-to-end training pipline, referenced in Fig.~\ref{fig:method}:
\begin{inparaenum}[\itshape i)\upshape]
\item SBS Sequentialization: We segment each $r_k$ into a sequence of action and state type $[(a_j,t_j)]_k$.
\item Fib Augmentation: We generate a batch of normal Fibonacci spheres for each $a_j$ and sampled a set of augmented ray $\mathbf{Fib}(r_k) = \{\hat{r_k}\}$. 
\item Decision Sequence Formulation: We apply $R_{\text{SBS}}$ to $\bm{R}_{tg}$ for the augmented dataset, shuffled with GT. We pad the wireless RT SBS with zero sequence to a uniform length and apply DT. 
\item State supervision ($\textsc{L}_{\text[type]}$): In addition to training against $\textsc{L}_{\text{ray}}(r_k,r'_k)$, we introduce additional set-based supervision on a type readout of each step $t'_i$ in $r'_k$, against GT interaction type $t_i$ attributed on respective state $s_i$. 
\end{inparaenum}

\section{Experiments}
\label{sec:exp}
We train SANDWICH and other baseline models for environments from the SOTA DeepLayout~\cite{Wu_DeepLayout_2019}. We create a standalone \textsc{gym} environment for each layout and test against $3$ test sets [\textsc{Checkerboard}, \textsc{genz}, \textsc{gendiag}],  as introduced in \cite{orekondy2023winert}. \textsc{Checkerboard} only includes OOD Tx sample, with identical Rx location, \textsc{genz} introduces Rx locates on a different horizontal plane, while \textsc{gendiag} introduces Rx location from a novel distribution in the 3D environment. All test sets include $15$ novel Tx locations and approximately $1800$ Rx locations, each (Tx, Rx) pair includes at most $30$ possible rays.~\footnote{\citet{orekondy2023winert} introduced $3$ new environments in "\textsc{wiIdoor}" besides the $100$ used environments in "\textsc{wi3room}". However, since its building block~\cite{remcomRemcomElectromagnetic} is proprietary software, we only use "\textsc{wi3room}" from DeepLayout~\cite{Wu_DeepLayout_2019} as baseline dataset.}

\subsection{Ablation Against Decision Transformer Families}
\begin{figure}
\centering\includegraphics[width=0.5\textwidth]{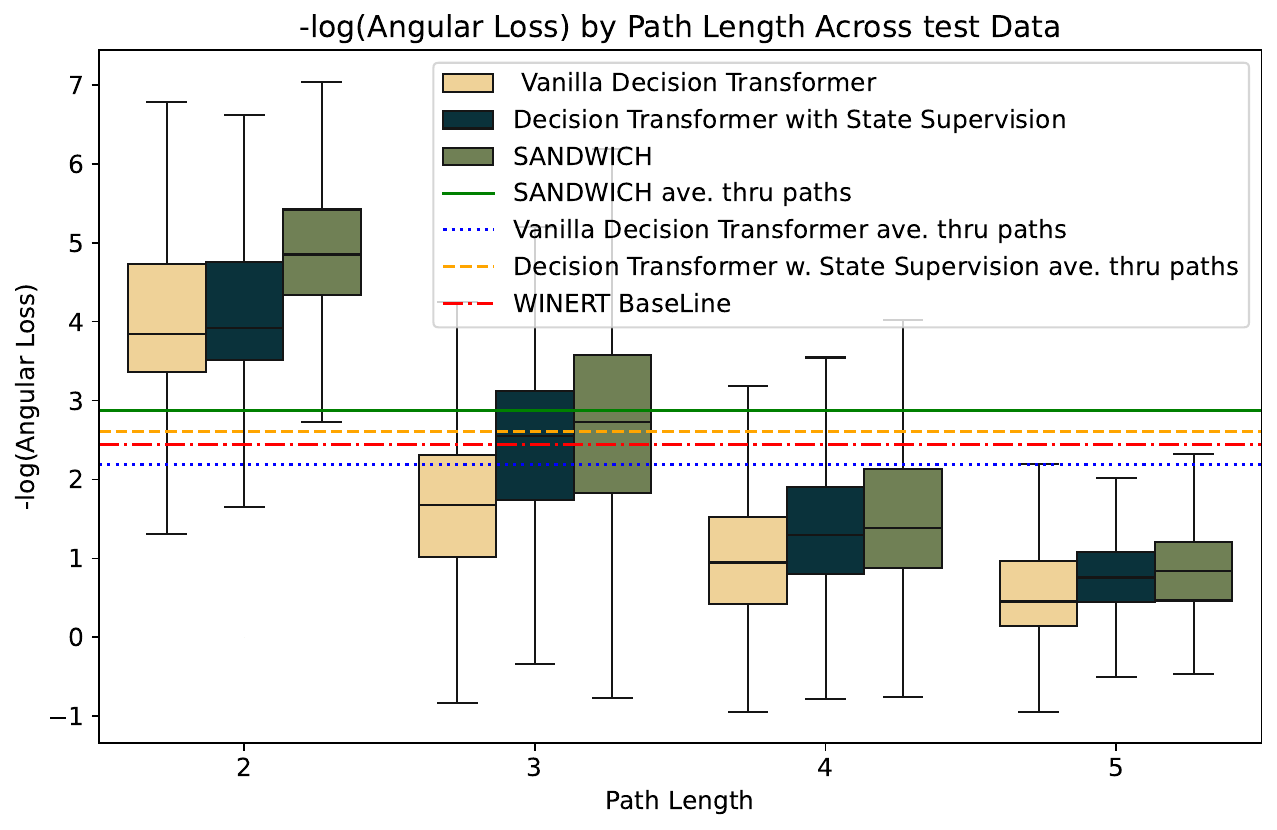}

\caption{Ablation of SANDWICH against Vanilia DT~\cite{chen2021decision} and DT with state supervision $\textsc{L}_{t}$~\cite{kimdynamics}.}    
    \label{fig:ablation}
\end{figure}

To prove the effectiveness of introduced novel building blocks, we conduct an ablation study against vanilla DT (train against $\textsc{L}_{\text{ray}}$) and DT with $\textsc{L}_{t}$. We train with the same hyper-parameter, test against all three test sets, and compare the best baseline from \cite{orekondy2023winert} as a reference in Fig.~\ref{fig:ablation}. Both the $\mathbf{Fib}(r_k)$ and $\textsc{L}_{\text[type]}$ contributes to SANDWICH's superior performance over vanilla DT and baselines. Also, due to bias propagation in generation task~\cite{ladhak2023pre}, applying $\textsc{L}_{\text[type]}$ and $\mathbf{Fib}(r_k)$ can restrain such degradation when path length increases.

\subsection{Wireless RT: Geometrical Accuracy}
We consider the performance of wireless RT accuracy by mean averaged angular loss in all three test sets. We compare our proposed solution against claimed and reproduced baselines in \cite{orekondy2023winert}. As no code was available for reproducing the baselines, for \textsc{MLP} baseline~\cite{sitzmann2020implicit,tancik2020fourier}, we consider a hyper-parameter search scheme towards downstream tasks for non-provided hyperparameters, i.e.: batch size, number of epochs, and formulation of angular supervision. For \textsc{KNN} baselines~\cite{sobehy2020csi}, we offer 2 forms of reproduction \begin{inparaenum} \item\label{knn:1} We reproduce according to \cite{orekondy2023winert} which offers an $n=1$ matching within Tx and Rx between train and test sets. \item\label{knn:2} We run $n=6$ KNN and pick the best-performing neighbor for angular loss, this upper-bounds the vanilla KNN reproduction performance as per claimed in~\cite{sobehy2020csi}. \end{inparaenum} For \textsc{WINERT} baseline, the logic component is not provided so we only apply claimed performance. We show that the proposed SANDWICH outperforms the previous baseline method in different generalizations of (Tx, Rx) locations. Although reported as SOTA, ~\cite{hehn2024differentiable} is not included in the comparison as they have not been open-sourced for either dataset or source code, making it impossible to ensure fairness.
\setcounter{footnote}{2}
\begin{table}[H]
\centering
\caption{Comparison of Wireless RT Angular Accuracy}
\begin{tabular}{lS[table-format=1.2]S[table-format=1.2]S[table-format=1.2]} 
\toprule
\textbf{Angular Loss (MAE, \unit{\radian})} & {Checkerboard} & {Genz} & {GenDiag}  \\ 
\midrule
SANDWICH (Proposed)~\textsuperscript{2}         & \bfseries 0.0464 & \bfseries 0.052  & \bfseries 0.0372 \\
WINERT (Claimed in \cite{orekondy2023winert})    & 0.087  & 0.084  & 0.085      \\
MLP(Claimed in \cite{orekondy2023winert})       & 0.33   & 0.35   & 0.322      \\
KNN(Claimed in \cite{orekondy2023winert})       & 0.212  & 0.226  & 0.213      \\
MLP(Reproduced)    & 0.504  & 0.518  & 0.497        \\
KNN(Reproduced~\ref{knn:1})    & 0.509    & 0.545    & 0.275        \\
KNN(Reproduced~\ref{knn:2}) & 0.470    & 0.512    & 0.247       \\
\bottomrule
\end{tabular}
\label{tab:angular}
\end{table}
\footnotetext{For $L_k=1$ ray is considered well received, as no learning model is applied.} 
\subsection{Turbo learning: RT as a prior for RSSI estimation}
Finally, we consider using generated rays as prior for received signal strength indicator (RSSI) prediction to examine the usability of wireless RT. The GCE serves as a topline that takes GT wireless RT as input, which bounds the best capability of the GCE. We apply the generated ray through SANDWICH to GCE and compare it against the topline and non-wireless RT-based solutions, with their implementations the same as in Table~\ref{tab:angular}. In Table~\ref{tab:mae_kl_comparison}, we show that SANDWICH generated ray provides similar performance as GCE using GT wireless RT result while outperforming other non-RT based baselines.


\section{Conclusion}
We propose SANDWICH, an offline, fully differentiable, GPU-trainable wireless RT neural surrogate. By using DT with state supervision and a novel data augmentation technique, we improve model accuracy. Our results show that SANDWICH provides wireless RT prior knowledge similar to a real raytracer in channel modeling. This work paves the way for efficient, reliable wireless RT solutions for JCAS integration.


\AtNextBibliography{\small}
\printbibliography

@inproceedings{orekondy2023winert,
  title={Winert: Towards neural ray tracing for wireless channel modelling and differentiable simulations},
  author={Orekondy, Tribhuvanesh and Kumar, Pratik and Kadambi, Shreya and Ye, Hao and Soriaga, Joseph and Behboodi, Arash},
  booktitle={The Eleventh International Conference on Learning Representations},
  year={2023}
}

@inproceedings{zhao2023nerf2,
  title={Nerf2: Neural radio-frequency radiance fields},
  author={Zhao, Xiaopeng and An, Zhenlin and Pan, Qingrui and Yang, Lei},
  booktitle={Proceedings of the 29th Annual International Conference on Mobile Computing and Networking},
  pages={1--15},
  year={2023}
}

@inproceedings{amiot2013pylayers,
  title={Pylayers: An open source dynamic simulator for indoor propagation and localization},
  author={Amiot, Nicolas and Laaraiedh, Mohamed and Uguen, Bernard},
  booktitle={2013 IEEE ICC},
  pages={84--88},
  organization={IEEE}
}

@inproceedings{hoydis2023sionna,
  title={Sionna RT: Differentiable ray tracing for radio propagation modeling},
  author={Hoydis, Jakob and Aoudia, Fay{\c{c}}al A{\"\i}t and Cammerer, Sebastian and Nimier-David, Merlin and Binder, Nikolaus and Marcus, Guillermo and Keller, Alexander},
  booktitle={2023 IEEE Globecom Workshops (GC Wkshps)},
  pages={317--321},
  organization={IEEE}
}

@article{choi2023withray,
  title={WiThRay: A versatile ray-tracing simulator for smart wireless environments},
  author={Choi, Hyuckjin and Oh, Jaeky and Chung, Jaehoon and Alexandropoulos, George C and Choi, Junil},
  journal={IEEE Access},
  volume={11},
  pages={56822--56845},
  year={2023},
  publisher={IEEE}
}

@misc{remcomRemcomElectromagnetic,
	author = {},
	title = {{R}emcom - {E}lectromagnetic {S}imulation {S}oftware},
	howpublished = {\url{https://www.remcom.com/}},
	year = {},
	note = {[Accessed 15-08-2024]},
}

@inproceedings{lai2024real,
  title={Real-Time Simulation of Wireless Signal Propagation for Dynamic Environment Through GPU-Based Ray Tracing},
  author={Lai, Zhongzheng and Zhang, Chenghao and Zhang, Yu and Yuan, Dong},
  booktitle={Proceedings of the ACM SIGCOMM 2024 Conference: Posters and Demos},
  pages={98--100},
}

@inproceedings{zhang2024wisegrt,
  title={Wisegrt: Dataset for site-specific indoor radio propagation modeling with 3d segmentation and differentiable ray-tracing},
  author={Zhang, Lihao and Sun, Haijian and Sun, Jin and Hu, Rose Qingyang},
  booktitle={2024 International Conference on Computing, Networking and Communications (ICNC)},
  pages={744--748},
  organization={IEEE}
}

@article{geok2018comprehensive,
  title={A comprehensive review of efficient ray-tracing techniques for wireless communication},
  author={Geok, Tan Kim and Hossain, F and Kamaruddin, MN and Rahman, NZA and Thiagarajah, Sharlene and Chiat, Alan Tan Wee and Liew, CP},
  journal={International Journal on Communications Antenna and Propagation},
  volume={8},
  number={2},
  pages={123--136},
  year={2018}
}

@software{instant_rm,
    title = {Instant Radio Maps},
    author = {Fayçal Aït Aoudia and Jakob Hoydis and Merlin Nimier-David and Sebastian Cammerer and Alexander Keller},
    note = {https://github.com/NVlabs/instant-rm},
    year = 2024
}

@inproceedings{ladhak2023pre,
  title={When do pre-training biases propagate to downstream tasks? a case study in text summarization},
  author={Ladhak, Faisal and Durmus, Esin and Suzgun, Mirac and Zhang, Tianyi and Jurafsky, Dan and McKeown, Kathleen and Hashimoto, Tatsunori B},
  booktitle={Proceedings of the 17th Conference of the European Chapter of the Association for Computational Linguistics},
  pages={3206--3219},
  year={2023}
}

@inproceedings{tarneberg2023towards,
  title={Towards Practical Cell-Free 6G Network Deployments: An Open-Source End-to-End Ray Tracing Simulator},
  author={T{\"a}rneberg, William and Fedorov, Aleksei and Callebaut, Gilles and Van der Perre, Liesbet and Fitzgerald, Emma},
  booktitle={2023 57th Asilomar Conference on Signals, Systems, and Computers},
  pages={1000--1005},
  organization={IEEE}
}

@article{chen2021decision,
  title={Decision transformer: Reinforcement learning via sequence modeling},
  author={Chen, Lili and Lu, Kevin and Rajeswaran, Aravind and Lee, Kimin and Grover, Aditya and Laskin, Misha and Abbeel, Pieter and Srinivas, Aravind and Mordatch, Igor},
  journal={NeurIPS},
  volume={34},
  pages={15084--15097},
  year={2021}
}

@article{hoydis2023learning,
  title={Learning radio environments by differentiable ray tracing},
  author={Hoydis, Jakob and Aoudia, Fay{\c{c}}al A{\"\i}t and Cammerer, Sebastian and Euchner, Florian and Nimier-David, Merlin and Ten Brink, Stephan and Keller, Alexander},
  journal={IEEE Transactions on Machine Learning in Communications and Networking},
  year={2024},
  publisher={IEEE}
}

@article{wang2024graph,
  title={Graph Neural Network enabled Propagation Graph Method for Channel Modeling},
  author={Wang, Xiping and Guan, Ke and He, Danping and Hrovat, Andrej and Liu, Ruiqi and Zhong, Zhangdui and Al-Dulaimi, Anwer and Yu, Keping},
  journal={IEEE Transactions on Vehicular Technology},
  year={2024},
  publisher={IEEE}
}

@article{knodt2021neural,
  title={Neural ray-tracing: Learning surfaces and reflectance for relighting and view synthesis},
  author={Knodt, Julian and Bartusek, Joe and Baek, Seung-Hwan and Heide, Felix},
  journal={arXiv preprint arXiv:2104.13562},
  year={2021}
}

@article{tancik2020fourier,
  title={Fourier features let networks learn high frequency functions in low dimensional domains},
  author={Tancik, Matthew and Srinivasan, Pratul and Mildenhall, Ben and Fridovich-Keil, Sara and Raghavan, Nithin and Singhal, Utkarsh and Ramamoorthi, Ravi and Barron, Jonathan and Ng, Ren},
  journal={NeurIPS},
  volume={33},
  pages={7537--7547},
  year={2020}
}

@article{sitzmann2020implicit,
  title={Implicit neural representations with periodic activation functions},
  author={Sitzmann, Vincent and Martel, Julien and Bergman, Alexander and Lindell, David and Wetzstein, Gordon},
  journal={NeurIPS},
  volume={33},
  pages={7462--7473},
  year={2020}
}

@article{samimi20163,
  title={3-D millimeter-wave statistical channel model for 5G wireless system design},
  author={Samimi, Mathew K and Rappaport, Theodore S},
  journal={IEEE Transactions on Microwave Theory and Techniques},
  volume={64},
  number={7},
  pages={2207--2225},
  year={2016},
  publisher={IEEE}
}

@article{paster2022you,
  title={You can’t count on luck: Why decision transformers and rvs fail in stochastic environments},
  author={Paster, Keiran and McIlraith, Sheila and Ba, Jimmy},
  journal={NeurIPS},
  volume={35},
  pages={38966--38979},
  year={2022}
}

@inproceedings{kimdynamics,
  title={Dynamics-Augmented Decision Transformer for Offline Dynamics Generalization},
  author={Kim, Changyeon and Kim, Junsu and Seo, Younggyo and Lee, Kimin and Lee, Honglak and Shin, Jinwoo},
  booktitle={3rd Offline RL Workshop: Offline RL as a''Launchpad''}
}

@article {Wu_DeepLayout_2019,
title = {Data-driven Interior Plan Generation for Residential Buildings},
author = {Wenming Wu and Xiao-Ming Fu and Rui Tang and Yuhan Wang and Yu-Hao Qi and Ligang Liu},
journal = {ACM Transactions on Graphics (SIGGRAPH Asia)},
volume = {38},
number = {6},
year = {2019},
}

@inproceedings{sobehy2020csi,
  title={CSI-MIMO: K-nearest neighbor applied to indoor localization},
  author={Sobehy, Abdallah and Renault, {\'E}ric and M{\"u}hlethaler, Paul},
  booktitle={2020 ICC (ICC)},
  pages={1--6},
  organization={IEEE}
}

@article{zhu2024physics,
  title={Physics-informed Generalizable Wireless Channel Modeling with Segmentation and Deep Learning: Fundamentals, Methodologies, and Challenges},
  author={Zhu, Ethan and Sun, Haijian and Ji, Mingyue},
  journal={arXiv preprint arXiv:2401.01288},
  year={2024}
}

@article{vaswani2017attention,
  title={Attention is all you need},
  author={Vaswani, A},
  journal={NeurIPS},
  year={2017}
}

@inproceedings{papaioannou2022integrated,
  title={Integrated ray-tracing and coverage planning control using reinforcement learning},
  author={Papaioannou, Savvas and Kolios, Panayiotis and Theocharides, Theocharis and Panayiotou, Christos G and Polycarpou, Marios M},
  booktitle={2022 IEEE 61st Conference on Decision and Control (CDC)},
  pages={7200--7207},
  organization={IEEE}
}

@inproceedings{ma2023rethinking,
  title={Rethinking decision transformer via hierarchical reinforcement learning},
  author={Ma, Yi and Jianye, HAO and Liang, Hebin and Xiao, Chenjun},
  booktitle={ICML},
  year={2023}
}

@misc{rogne2022raytracing,
  title={Raytracing in channel model development},
  author={Rogne, Andreas},
  year={2022}
}

@misc{rtem, 
    title = {Advanced electromagnetic ray tracing methods},
    howpublished = {\url{www.ee.cit.tum.de/en/hft/forschung/advanced-electromagnetic-ray-tracing-methods/}},
    journal={Chair of High-Frequency Engineering},
    publisher={TUM School of Computation, Information and Technology},
    note = {[Accessed 15-08-2024]}}

@ARTICLE{18706,
  author={Ling, H. and Chou, R.-C. and Lee, S.-W.},
  journal={IEEE Transactions on Antennas and Propagation}, 
  title={Shooting and bouncing rays: calculating the RCS of an arbitrarily shaped cavity}, 
  year={1989},
  volume={37},
  number={2},
  pages={194-205},
  doi={10.1109/8.18706}}

@article{hehn2024differentiable,
  title={Differentiable and Learnable Wireless Simulation with Geometric Transformers},
  author={Hehn, Thomas and Peschl, Markus and Orekondy, Tribhuvanesh and Behboodi, Arash and Brehmer, Johann},
  journal={arXiv preprint arXiv:2406.14995},
  year={2024}
}

@inproceedings{chi2024rf,
  title={RF-Diffusion: Radio Signal Generation via Time-Frequency Diffusion},
  author={Chi, Guoxuan and Yang, Zheng and Wu, Chenshu and Xu, Jingao and Gao, Yuchong and Liu, Yunhao and Han, Tony Xiao},
  booktitle={Proceedings of the 30th Annual International Conference on Mobile Computing and Networking},
  pages={77--92},
  year={2024}
}

@inproceedings{carrpath,
  title={PATH TRACING: A NON-BIASED SOLUTION TO THE RENDERING EQUATION},
  author={Carr, Robert and Hulcher, Byron},
 booktitle={Advanced Graphics Lecture Notes},
  year={2011}
}

@inproceedings{clarberg2022keynote,
  title={Are we done with ray tracing?},
  author={Keller, Alexander and Viitanen, Timo and Barr{\'e}-Brisebois, Colin and Schied, Christoph and McGuire, Morgan},
  booktitle={SIGGRAPH Courses},
  pages={3--1},
  year={2019}
}

@inproceedings{kajiya1986rendering,
  title={The rendering equation},
  author={Kajiya, James T},
  booktitle={Proceedings of the 13th annual conference on Computer graphics and interactive techniques},
  pages={143--150},
  year={1986}
}

@article{radford2018improving,
  title={Improving language understanding by generative pre-training},
  author={Radford, Alec},
  year={2018}
}

@inproceedings{feng2024d,
  title={D-Band Channel Modelling by 3D Ray Tracing for Joint Communications and Sensing},
  author={Feng, Yunqi and Guenach, Mamoun and Bourdoux, Andr{\'e} and Joseph, Wout and Tanghe, Emmeric},
  booktitle={2024 IEEE 4th International Symposium on Joint Communications \& Sensing (JC\&S)},
  pages={1--6},
  year={2024},
  organization={IEEE}
}
\balance
\end{document}